\documentclass{aa}
\usepackage{graphicx}
\usepackage{txfonts}

\begin{document}
\title{The CFHTLS Real Time Analysis System:\\
"Optically Selected GRB Afterglows"\thanks{Based on 
observations obtained with MegaPrime/MegaCam, a 
joint project of CFHT and CEA/DAPNIA, at the Canada-France-Hawaii 
Telescope (CFHT) which is operated by the National Research Council 
(NRC) of Canada, the Institut National des Science de l'Univers 
of the Centre National de la Recherche Scientifique (CNRS) of 
France, and the University of Hawaii.}}
\subtitle{I. Overview and performance}

\author{F. Malacrino\inst{1}
       \and
       J-L. Atteia\inst{1}
       \and
       M. Bo\"er\inst{2}
       \and
       A. Klotz\inst{2}$^{,}$\inst{3}
       \and
       C. Veillet\inst{4}
       \and
       J-C. Cuillandre\inst{4} on behalf of the GRB RTAS Collaboration}
\offprints{F. Malacrino, fmalacri@ast.obs-mip.fr}

\institute{Laboratoire d'Astrophysique de Toulouse-Tarbes, Observatoire Midi-Pyr\'en\'ees 
(CNRS-UMR5572/Universit\'e Paul Sabatier Toulouse III), 14 Avenue Edouard Belin, 31400 Toulouse, France\\
          \email{fmalacri@ast.obs-mip.fr}
          \and Observatoire de Haute-Provence, 04870 Saint-Michel l'Observatoire, France
          \and Centre d'Etude Spatiale des Rayonnements, Observatoire Midi-Pyr\'en\'ees 
(CNRS/UPS), BP 4346, 31028 Toulouse Cedex 04, France
	  \and Canada-France-Hawaii Telescope Corp., Kamuela, HI 96743, USA
         }

\date{Received \today; accepted}

\abstract
{}
{We describe a deep and wide search for optical GRB afterglows in images 
taken with MegaCAM at the Canada France Hawaii Telescope, within the
framework of the CFHT Legacy Survey.}
{This search is performed in near real-time thanks to a Real Time 
Analysis System (RTAS) called "Optically Selected GRB Afterglows", 
which has been installed on a dedicated computer in Hawaii. 
This pipeline automatically and quickly analyzes 
Megacam images to construct catalogs of astronomical objects, and
compares catalogs made from images taken at different
epochs to produce a list of astrometrically and 
photometrically variable objects. These objects are then displayed on a web page 
to be characterized by a member of the collaboration.}
{In this paper, we comprehensively describe the RTAS process from 
image acquisition to the final characterization of variable sources. 
We present statistical results based on one full year 
of operation, showing the quality of the images and the performance 
of the RTAS. The limiting magnitude of our search is r'=22.5 on average
and the observed area amounts to 1178 square degrees. We have 
detected about 13.10$^6$ astronomical sources of which about 
0.01\% are found to vary by more than one tenth of a magnitude.
We discuss the performance of our instrumental setup
with a sample of simulated afterglows. This sample
allows us to compare the efficiency of our search with previous works, 
to propose an optimal observational strategy, 
and to discuss general considerations on the searches 
for GRB optical afterglows.
We postpone to a forthcoming paper the discussion of the characterization 
of variable objects we have found, and a more detailled analysis of the nature of 
those resembling GRB afterglows.}
{The RTAS has been continuously operating since November 2004. 
Each month 15-30 square degrees are observed many times over a period 
of 2-3 nights. The real-time analysis of the data has revealed no 
convincing afterglow candidate so far.}


\maketitle


\section{Introduction}
Long gamma-ray bursts (hereafter GRBs) are cosmological events
due to powerful stellar explosions in distant galaxies. 
They are composed of two phases: the prompt emission, a short 
and bright flash of $\gamma$-ray and X-ray photons, and the afterglow, 
a fainter decaying emission visible from X-ray to radio wavelengths. 
Current observations of the prompt GRB emission and of the afterglow are 
satisfactorily described in the framework of the internal-external 
shock model. This model explains the prompt emission as the radiation
emitted by internal shocks within an unsteady outflow of ultra-relativistic
material (Rees \& M{\'e}sz{\'a}ros \cite {rees94}), and the afterglow by the shock 
of the ultra-relativistic outflow on the medium surrounding the source
(e.g. Rees \& M{\'e}sz{\'a}ros \cite {rees92}, M{\'e}sz{\'a}ros \& Rees \cite {meszaros97},
Wijers et al. \cite {wijers97}).
Based on theoretical and observational grounds, there is now a general 
consensus on the fact that the prompt GRB emission is collimated into a jet
which broadens gradually as its bulk Lorentz factor decreases.

One strong argument in favor of GRB collimation is that it greatly reduces 
the energy requirements at the source. For example an event like GRB 990123
would have released of the order of 3 10$^{54}$ erg of high-energy radiation 
if it were radiating isotropically (Kulkarni et al. \cite {kulkarni99}). 
This energy budget can be reduced to 2 10$^{51}$ erg if we assume that 
the prompt $\gamma$-ray emission was collimated into two opposite jets 
with a FWHM of 2.9$^\circ$ (e.g. Frail et al. \cite{frail01}).
Theoretical calculations of the evolution of the ejecta have shown
that the afterglows of beamed gamma-ray bursts must exhibit 
an achromatic 'jet break' when $1/\Gamma$, 
the inverse of the bulk Lorentz factor of the jet, becomes comparable to $\theta$, 
the opening angle of the jet (Rhoads \cite{rhoads97}). From the observational point of view, the light-curves of several GRB 
afterglows display achromatic breaks, observed at optical and X-ray wavelengths,
hours to days after the burst (e.g. Harrison et al.  \cite{harrison99}),
providing strong observational evidence in favor of GRB beaming.  
Interpreting these breaks as signatures of the beaming of the high-energy emission 
gives opening angles ranging from 3\degr\ to 30\degr. 
Even if the signature of GRB jets in the light-curves of the afterglows 
remains a subject of debate
(see for instance Wei \& Lu \cite{wei00}, Moderski et al. \cite{moderski00}) and if other causes can produce breaks in GRB afterglow light-curves 
(e.g. cooling breaks, Sari \cite {sari98}),
the jet break interpretation is supported by the intriguing fact
that the energy output corrected from beaming appears well clustered 
around 10$^{51}$ erg, with a dispersion much smaller than the energy 
output obtained assuming isotropic emission
(Frail et al. \cite {frail01}, Panaitescu \& Kumar \cite {panaitescu01},
Bloom, Frail \& Kulkarni \cite {bloom03}). 
This result can be interpreted as the evidence that GRBs have a
standard energy reservoir (Frail et al. \cite {frail01}, Panaitescu \& Kumar 
\cite {panaitescu01}, Piran et al.  \cite {piran01}), or as the evidence
that GRB jets have a universal configuration (Zhang \& M{\'e}sz{\'a}ros \cite {zhang02}).

One remarkable prediction of the models of jetted GRBs is that the jet 
starts to spread out when the afterglow is still detectable, 
allowing the afterglow to become visible for off-axis observers. 
This prediction has led to the concept of 'orphan afterglows',
initially used for the afterglows of off-axis GRBs. 
Afterglows of off-axis GRBs have been studied from a theoretical 
point of view by many authors (Rhoads \cite {rhoads97}, Rhoads \cite{rhoads99}, 
Wei \& Lu \cite {wei00}, Totani \& Panaitescu \cite {totani02}, 
Nakar et al. \cite {nakar02}, Dalal et al. \cite{dalal02}).
Recently, it has been realized that orphan afterglows could be also produced
by failed on-axis GRBs, which are fireballs with Lorentz factors 
well below 100 but larger than a few (Huang et al. \cite{huang02}).

This short discussion illustrates the reasons that make the existence 
of orphan afterglows quite probable (both from off-axis GRBs and 
from failed GRBs), and their relation with GRBs non-trivial
(e.g. Dalal et al. \cite{dalal02}, Huang et al. \cite{huang02}).
Given the potential pay-off which would result from the detection
of even a few orphan afterglows (energetics, distance, rate of 
occurrence...), it is important trying detecting them.
As explained above, the detection of orphan GRB afterglows 
offers a complementary way to test the beaming hypothesis and 
can constrain the beaming factor. 
Additionally, since afterglows of off-axis GRBs are expected to be 
more numerous but fainter, they will be detectable at lower redshifts.
Their detection could thus help estimating the local population 
of faint GRBs ($z<<0.1$), of which GRB980425 (z=0.0085) and 
GRB060218 (z=0.033) are the best known examples. 

In short, we are motivated by the fact  
that the detection of orphan afterglows may open a completely 
new way to detect GRBs and permit the study of 
a population of GRBs which is not or very poorly studied at present
(all GRBs known to date having been detected by their high-energy emission). 
The motivations driving the searches for GRB orphan 
afterglows have been discussed by various authors like 
Totani \& Panaitescu (\cite{totani02}), 
Nakar \& Piran (\cite{nakar02}), 
Kehoe et al. (\cite{kehoe02}), 
Groot et al. (\cite{groot03}), 
Rykoff et al. (\cite{rykoff05}) in the optical range,
by Greiner et al. (\cite{greiner99}) for X-ray afterglows, and
by Perna \& Loeb (\cite{perna98}), 
Levinson et al. (\cite{levinson02}), 
Gal-Yam et al. (\cite{galyam06}) for radio afterglows.
One difficulty of this task, however, is that we have little theoretical
indication on the rate and luminosity of orphan afterglows, two parameters 
which are essential in designing a strategy to search these sources.
The scarcity of the GRBs suggests nevertheless that the detection of 
orphan afterglows will require the monitoring of a wide area of the sky.

\begin{figure}[ht]
\centering
\resizebox{\hsize}{!}{\includegraphics{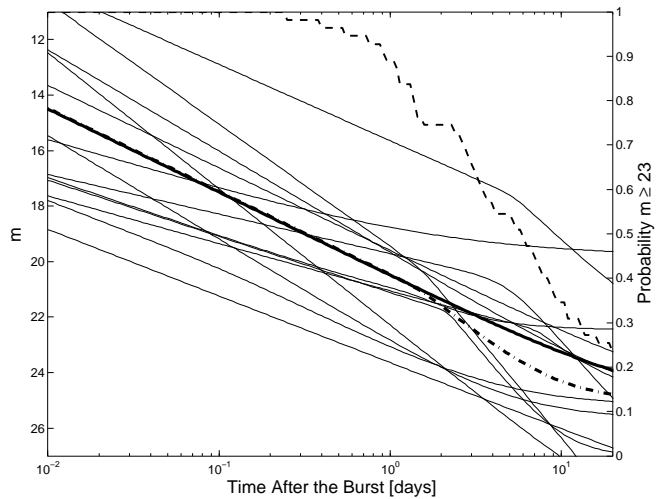}}
\caption{This figure shows 14 afterglow light-curves generated with 
parameters summarized in Zeh, Klose \& Kann (\cite{zeh}). 
The bold line and the bold dash-dotted line show the light curves of 
a typical afterglow ($\alpha_{1}=1,2\,$, $\alpha_{2}=2$, $t_b=1,5$ days,
 $m_1=20,5$, $m_b=21$, $m_h=25$) without and with a break respectively.
 The bold dashed line is the fraction of afterglows brighter than 
$23^{rd}$ magnitude at a given time (right hand scale). Nearly 90\% 
of the afterglows are visible at magnitude $m\leqslant23$, 
one day after the burst and 50\%, 6 days after the burst.}
\label{allag}
\end{figure}

Untriggered searches for GRB afterglows have already been attempted 
by a few teams, with really different observational parameters, 
but unsuccessfully. Rykoff et al. (\cite {rykoff05}) performed such 
a search with the ROTSE-III telescopes, covering a wide field at 
low sensitivity. On the contrary, Becker et al. (\cite {becker}) 
have favoured a very deep survey, but with a very small field of view. 
They have found two interesting transient objects, which have recently 
been confirmed as flare stars (Kulkarni \& Rau \cite {kulkarni06}). 
A similar attempt has also been performed by Rau et al. (\cite {rau}), 
and once again an object behaving like a GRB afterglow has been 
detected, which was later identified as a flare star (Kulkarni \& Rau \cite {kulkarni06}). 
Vanden Berk et al. (\cite {van}) have 
searched GRB afterglows in the data of the SDSS survey. 
A very interesting transient was found, but it appeared to be 
an unusual AGN (Gal-Yam et al. \cite {galyam02}). So far no convincing 
optically selected GRB afterglow has been found. The failure of these 
searches is essentially the consequence of the scarcity of GRBs and of 
the faintness of their afterglows. The combination of these two
factors implies that searches for orphan afterglows must be deep
and cover several percent of the sky 
to have a reasonable chance of success.
The search presented in this paper has a  magnitude limit which is 
about the same as the search of Rau et al. (\cite {rau}), but a 
sky coverage which is 50 times larger.

Our search for untriggered GRB afterglows uses images collected for the Very Wide Survey (one of
the three components of the CFHT Legacy Survey, see http://www.cfht.hawaii.edu/Science/CFHTLS/) at the 
Canada-France-Hawaii Telescope. 
In the next section, 
we present the observational strategy used in this study. 
Section 3 introduces the Real Time Analysis System, with more details 
given in section 4 (catalog creation) and in section 5 (catalog comparison). 
In section 6 
we analyze the performance of the RTAS during one full year 
of observation. Comparison with previous studies will be done in section 7. 
The last two sections encompass global considerations on afterglow search, the description 
of an 'optimal' survey and our conclusions. All the web pages mentioned in this paper can be found at 
our web site: http://www.cfht.hawaii.edu/$\sim$grb/

\section{The Canada-France-Hawaii Legacy Survey}
The Canada-France-Hawaii Telescope (hereafter CFHT) is a 3.6 meter telescope 
located on the Mauna Kea in the Big Island of Hawaii. Built in the late 70's, 
it has been equiped in 2003 with  a high-performance instrument, MegaCAM. 
MegaCAM is a 36 CCD imager covering about 1 square degree field of view. 
Each CCD frame has $2\,048\times4\,612$ pixels, for a total of $340\times10^{6}$ pixels. 
It observes the sky through 5 filters (u* g' r' i' z'), with a resolution 
of 0.185\arcsec per pixel. These characteristics, combined with the excellent 
climatic conditions at the site, provide very good quality images.

\begin{figure}[ht]
\centering
\resizebox{\hsize}{!}{\includegraphics{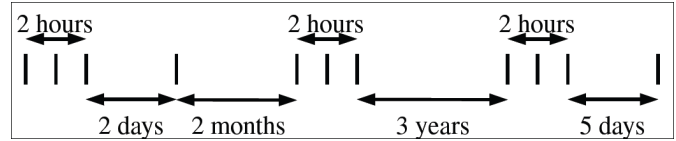}}
   \caption{This diagram shows the observational strategy for one field of 
the Very Wide Survey. Each vertical line stands for one exposure, the exposure 
time depends on the filter, but is typically of the order of 100 seconds. 
15 new fields or more are observed each month.
   }
   \label{vw}
\end{figure}

The CFHT Legacy Survey is the main observing program at the CFHT since june 2003. 
It is composed of 3 different surveys:
\begin{itemize}
\item[-] The Wide Synoptic Survey, covering 170\,deg$^{2}$ with all the 5 
MegaCAM filters (u* g' r' i' z') down to approximatively i'=25.5. 
The main goal of this survey is to study large scale structure and 
matter distribution in the universe.
\item[-] The Deep Synoptic Survey which covers 4\,deg$^{2}$ down to i'=28.4, 
and through the whole filter set. Aimed mainly at the detection of 2\,000 
type I supernov\ae{} and the study of galaxy distribution, this survey will 
allow an accurate determination of cosmological parameters.
\item[-] The Very Wide Survey, covering 1\,200\,deg$^{2}$ down to i'=23.5, 
with only 3 filters (g' r' i'). As it has been initially conceived to discover 
and follow Kuiper Belt Objects, each field is observed several times, according 
to the strategy explained in Fig.~\ref{vw}.
\end{itemize}

The images taken by the CFHT are pre-processed by a pipeline called Elixir 
(Magnier \& Cuillandre \cite{magnier}), which flattens and defringes each 
CCD frame and computes gross astrometry and photometry. About 20 minutes are needed 
to transfer the data from the telescope to Waimea and to process them with Elixir. 
Thanks to this pipeline, calibrated images are 
available in quasi-real time for the RTAS.

Although the Deep Synoptic Survey has a very interesting observational 
strategy, preliminary simulations have shown that the number 
of afterglow detections expected in near real time is very low compared to the Very Wide Survey.
With relatively deep observations and very good quality images, 
the Very Wide Survey represents a credible opportunity to detect 
GRB afterglows independently of the prompt emission. 
Moreover, this is the only sub-survey with a well defined observational 
recurrence which can be used to compare images between them in order 
to detect variable, new, and/or vanishing objects, such as GRB afterglows. 
However, since we restrict the comparisons to objects detected in images taken 
during the same run, we are only able to detect objects with strong 
and fast variability.

\section{The Real Time Analysis System}
It is generally accepted that the most important quality in afterglow 
detection is speed. The Real Time Analysis System 
has been built to allow a quick 
follow-up of the afterglow candidates. Its goal is to analyze 
in quasi-real time ($<24$ hours) images of the Very Wide Survey to detect objects 
behaving like afterglows of GRBs. To permit quick automatic analysis, 
we have decided to work with catalogs of objects, and to compare between 
them catalogs of the same field of the sky taken at different times. 
Although the USNO-A2 catalog (Monet et al. \cite{monet98})
is used to astrometrically calibrate images, and the Digitized Sky 
Survey (DSS) to quickly characterize bright objects, we have no reference 
but ourself for objects fainter than $\sim20$ magnitude, because 
of the depth and width of the Very Wide Survey.

Observations along the year at the CFHT are divided into runs, which are periods 
between two full moons lasting about 2 weeks. The optimization of the CFHT observational 
strategy during a run implies that we don't know in advance when the images for the Very Wide Survey 
will be taken, because this depends on the weather, and on the pressure from 
other observational programs in queue observing mode. Two half-nights are 
generally dedicated to the Very Wide Survey, during which 15 fields at least
are observed once to four times. This unpredictability is the reason why 
it is very important for the RTAS to be fully automatic. The main script 
is launched every fifteen minutes to check for new images to be processed, 
and the whole process is done without human intervention.

The RTAS has been installed on a dedicated computer at the CFHT headquarters. It is composed of many Perl scripts which prepare files 
and generate code for the other software used by the process. 
FITS to GIF conversion is done with IRAF\footnote{IRAF is distributed 
by the National Optical Astronomy Observatories, which are operated by 
the Association of Universities for Research in Astronomy, Inc., under 
cooperative agreement with the National Science Foundation.}, catalogs 
of objects with SExtractor, and the system core is coded for Matlab. 
Perl scripts also generate HTML and CSS outputs, which become accessible 
from the CFHT web page. CGI scripts used for catalog and comparison 
validation are also coded in Perl.

\begin{figure}[ht]
\centering
\resizebox{\hsize}{!}{\includegraphics{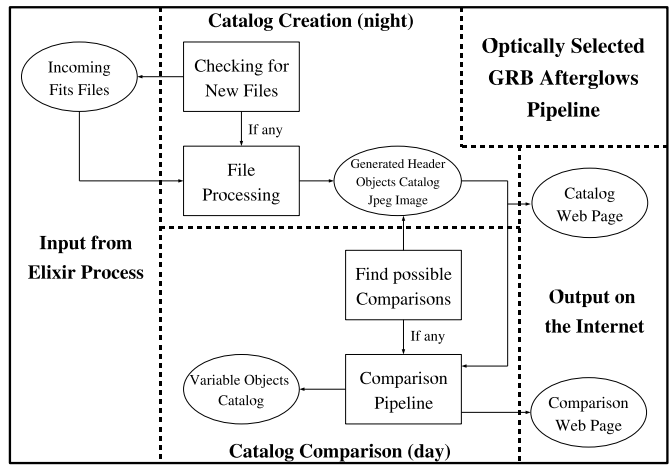}}
   \caption{This diagram shows the global mechanism of the RTAS 
pipeline and the interactions between the different components}
   \label{pipeline}
\end{figure}

As shown in Fig.~\ref{pipeline}, the RTAS can be divided into two 
distinct parts, a \emph{night process} for the catalog creation, and a \emph{day 
process} for the catalog comparison. All results generated by the automatic 
pipeline are summarized on dynamic HTML web pages. Members of the 
collaboration are then able to check the process with a nice interface 
from any place which has internet connectivity.

\section{Catalog Creation}
The catalog creation process is composed of two main parts. 
The first part consists in the reduction of the useful 
information from about 700 MB, the size of an uncompressed MegaCAM 
image processed by Elixir, to a few tens of MB\footnote{Dealing 
with smaller data sets makes the treatment much faster and guarantees the permanent availability of the whole survey database 
on a commercial machine with moderate disk space (about 300 GB)}. 
The second part prepares the comparison between catalogs by 
astrometrically and photometrically calibrating objects, and 
sorting them according to their astrophysical properties.

Although images from the Very Wide Survey are processed one by one, 
the entire process works separately with each of the 36 CCD frames of 
an image, for two main reasons. First, the Elixir pipeline outputs 
one FITS file per CCD frame, so it is easier to use the same kind of data. 
Second, the astrometric calibration, which is the most time consuming 
operation in the catalog creation process, is made faster, but still 
accurate enough, since each CCD frame contains from 500 to a few thousand 
objects, depending on the filter used, the exposure time and the pointed 
sky region (see Fig.~\ref{usnomag}). Working with individual CCD frames 
has no major impact on the RTAS. The comparisons can be done within each CCD frame separately, since the pointing of the telescope 
is highly reproducible, and the CCD frames in images taken 
at different times overlap almost exactly. 

In a first pass, the process detects the presence of new files in the 
directory where the Elixir pipeline pushes processed MegaCAM images, 
and waits for the image to be complete (with one file for each CCD frame). 
Images which are not part of 
the Very Wide Survey or which have already been processed are rejected; 
otherwise a backup of the files is done, allowing to reprocess them in case of an error in the 
treatment. In a second time, the FITS files of each CCD frame are converted 
into GIF images using IRAF, and then into JPEG images using the unix 
\textit{convert} command. Then, the FITS header for each CCD frame is extracted and 
copied in an ASCII file. Some entries contained in this header are 
pushed as input parameters for SExtractor, used to create the catalogs.

In particular, an aproximative magnitude zero point (Mag0Point) is 
computed using header information. We have to mention here that 
the value of this self-computed Mag0Point does not take into account 
the climatic conditions of observations, especially seeing conditions 
and extinction due to clouds. It means that magnitudes of objects are 
relative, not absolute, although most of time the value is very 
close to the real one. Then, SExtractor is launched  in order to 
create the catalog of objects, and the input parameters are added 
to the header as a reminder.

In parallel, an ASCII catalog of the same region of the sky is 
extracted from the USNO-A2.0 catalog, and used in an improved triangle 
matching method (Valdes et al. \cite {focas}) in order to astrometrically 
calibrate our catalog. We achieve a precision of 0.6\arcsec or better for 
each CCD frame for the absolute positions of objects (see Table.~\ref{sinstat}). 
We have decided to not photometrically match objects with the USNO-A2.0 
catalog, because the filters used do not correspond. At this step of 
the process, the catalogs contain one line per object, 
with the following parameters:
\begin{itemize}
\item[-] A unique ID number
\item[-] The pixel coordinates, X and Y
\item[-] The J2000 coordinates, Right Ascension and DEClination
\item[-] The $\mu_{max}$\footnote{The $\mu_{max}$ is the magnitude 
of the brightest pixel of the object} and the magnitude
\item[-] The FWHM (Full-Width at Half-Maximum)
\item[-] A flag computed by SExtractor: if its value doesn't 
equal 0, the values of parameters are not reliable
\end{itemize}

\begin{figure}[ht]
\centering
\resizebox{\hsize}{!}{\includegraphics{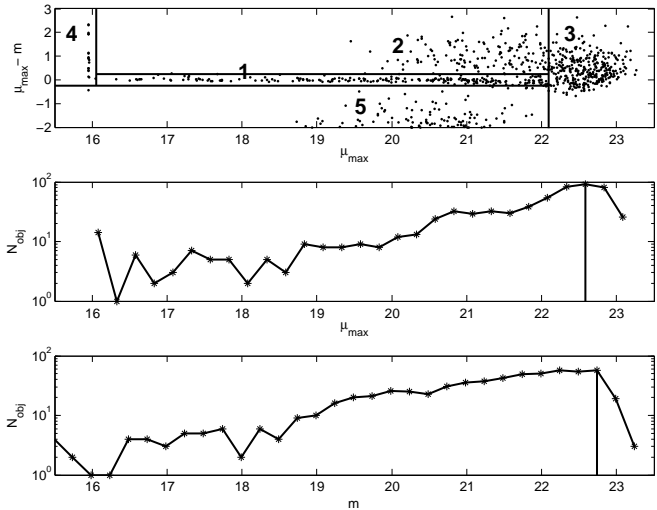}}
\caption{The upper figure shows our classification 
of objects in a plot $\mu_{max}-$magnitude versus $\mu_{max}$. 
We construct 4 classes of astrophysical objects: stars (1), galaxies (2), 
faint objects (3), saturated objects (4), and a class containing
cosmic-rays (5). The lower two graphics represent 
the computation of the $\mu_{max}$ and magnitude completeness respectively.
   }
   \label{sort}
\end{figure}

In a next step, objects are sorted according to their astrophysical properties. 
By comparing $\mu_{max}$, the equivalent magnitude of the brightest pixel 
of the object, and the magnitude, we are able to find the 
"line of stars"\footnote{Since stars are point sources $\mu_{max}$, the brightness
of their brightest pixel, is exactly proportional to their magnitude
when the image is well oversampled. 
On a given CCD frame the difference $\mu_{max}$-magnitude is constant for
stars, which follow a well defined 'line' in a plot showing $\mu_{max}$-magnitude
as a function of $\mu_{max}$. The position of this 'line' can change according
to the observing conditions.}, and so to separate objects into 5 classes, 
of which 4 are astrophysical objects: stars, galaxies, faint objects, satured objects, 
and the last one contains cosmic-rays (see Fig.~\ref{sort}). 
Although this classification is quite arbitrary, especially 
for the faint objects boundary, it is very useful to compare objects 
between them and to reject non-astrophysical ones. Finally, $\mu_{max}$ and 
magnitude completeness will be used as cuts in the comparison 
process (see Fig.~\ref{sort}) while the "line of stars" will be used 
to intercalibrate $\mu_{max}$ and magnitude between different images.

The processing of one image usually lasts between 5 and 15 minutes, 
mainly depending on the filter and of the observed region of the sky. 
The most time consuming steps are SExtractor and the astrometric matching. 
Most of the errors come from the USNO matching in CCD frames containing 
a very bright star or a large number of objects. These CCD frames are 
then flagged as unusable, and safe data are backed up on a special 
directory, allowing a quick re-processing of the image with the corrected 
code. Less than $0.35\%$ of the CCD frames produce an error (see table~\ref{sinstat}). 
Data of CCD frames correctly processed are saved in a database, which makes 
the post-process of the RTAS independent from the CFHT, and allows 
quick search of all kind of information in the whole set of data already processed.

Finally, all the results of the catalog creation process are summarized in real-time 
on an automatically generated HTML web page. 
In this page we summarize photometric, astrometric and classification 
values. Using an interactive script, collaboration members can check 
the results of the catalog creation process and decide to validate it. 
This validation allows starting the second part of the processing which involves 
the comparison of catalogs of the same field.

\section{Catalog Comparisons}
The goal of this process is to compare catalogs and extract from them 
a list of variable objects. The comparisons involve 
images of the same field, taken through the same filter. Exposure times 
have also to be of the same order. In a first step, catalogs are compared 
within a single night, by doublets (two images of the same field) 
or triplets (three images of the same field), depending on the Very Wide Survey observational strategy for this night. 
Differences between triple and double comparisons are discussed 
below. In a second step, the process selects for 
each field the best quality image of each night within the current run, 
and keeps it for an inter-night double comparison.

\subsection{Triple Comparisons}
Triple comparisons aim at extracting objects with strong magnitude 
variations, detecting asteroids and TNOs, and creating a reference catalog. 
Triple comparisons always involve images acquired during the same night.

\begin{figure}[ht]
\centering
\resizebox{\hsize}{!}{\includegraphics{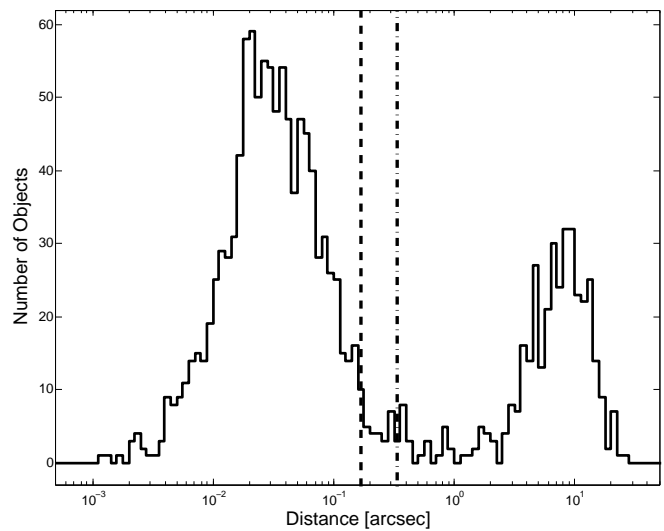}}
   \caption{Histogram of the distance separating the nearest objects
in two images. The largest peak is due to real astrophysical objects which
are detected in the two images. The two vertical lines show the position 
tolerance and two times the position tolerance used for the matching of objects
in the two images.}
   \label{distances}
\end{figure}

\begin{figure}[ht]
\centering
\resizebox{\hsize}{!}{\includegraphics{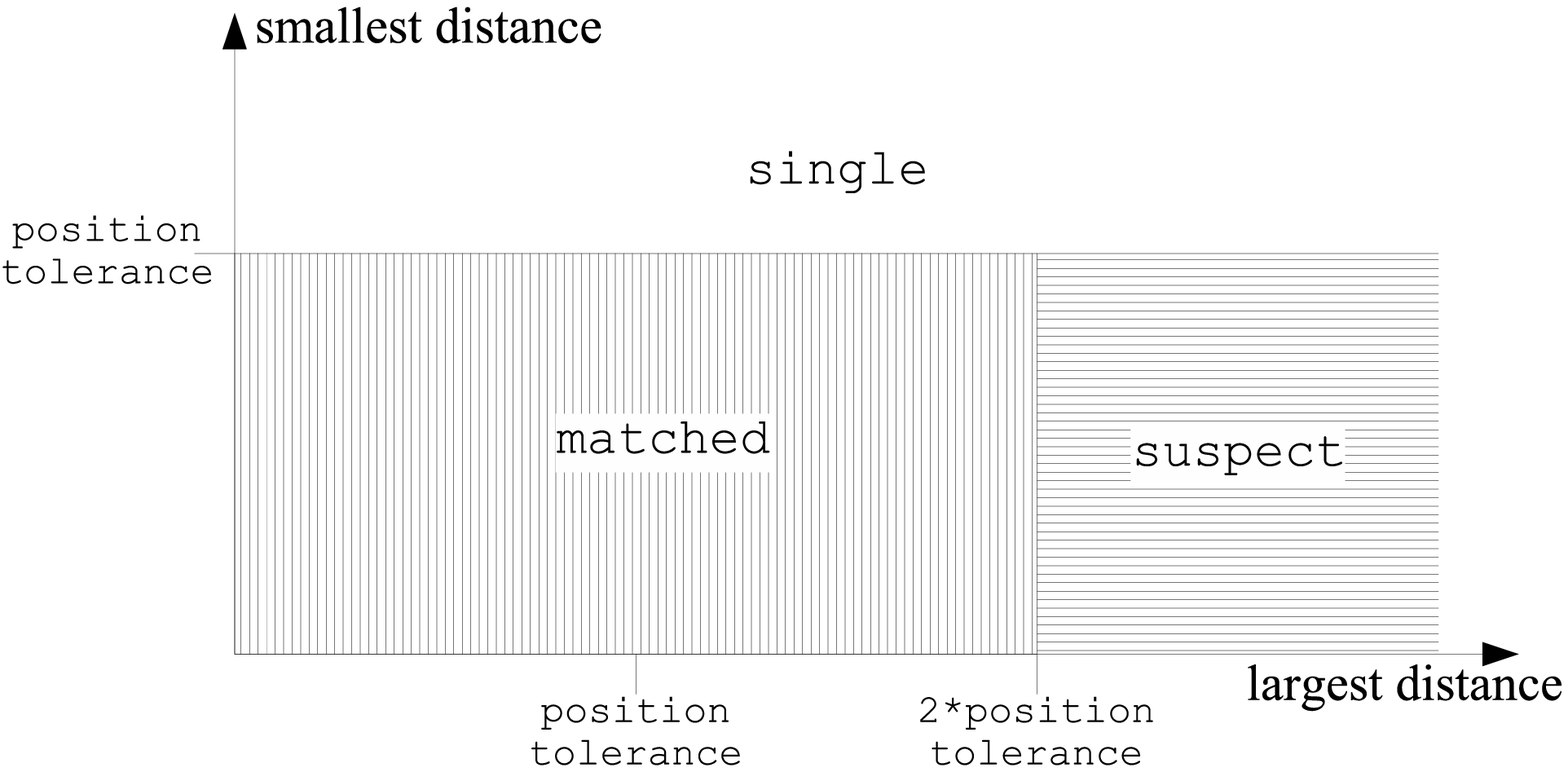}}
   \caption{This diagram illustrates the spliting of astrophysical objects in 
matched, suspect and single objects in triple comparisons}
   \label{carac}
\end{figure}

Before the beginning of the comparison, catalogs are ordered by ascending 
quality, defined by the mean number of astrophysical objects per CCD frame. 
The catalog with the highest quality (hereafter catalog 3) 
is taken as touchstone for the other two. 
All the calibrations are done with respect to this catalog. 
First, catalogs 1 and 2 are astrometrically matched to catalog 3 
using the same method as for the USNO matching in the catalog creation. After this step we compute for each object the distance to its nearest 
neighbour in the other two catalogs.
The nearest neighbour distance distribution is used to determine a 
position tolerance beyond which objects are considered to be distincts. 
This value is usually of the order of 1 pixel, or 0.2\arcsec (see Fig.~\ref{distances}).
The objects are then classified into 3 categories depending 
on their distances with their nearest neighbours in the other two catalogs, 
wich are compared with the position tolerance derived above :
\begin{itemize}
\item[-] If the smallest distance is lower than the position tolerance and 
the largest one is lower than twice the position tolerance, 
the object is classified as \textbf{matched}.
\item[-] If the smallest distance is lower than the position tolerance and 
the largest distance is higher than twice the position tolerance, 
the object is classified as \textbf{suspect}.
\item[-] Otherwise, the object is classified as \textbf{single}.
\end{itemize}
To summarize, matched objects are in all catalogs, suspect objects in two 
of them, and single objects are in only one catalog. A visual representation 
of this classification based on spatial proximity can be found in Fig.~\ref{carac}.

Once all objects have been classified, matched stars are used to calibrate 
magnitudes, $\mu_{max}$ and FWHM to catalog 3. 
Matched objects which have been classified as stars, galaxies or faint 
objects in the catalog creation process are then searched for variability. 
The pipeline for the detection of variable objects works with 9 magnitude 
bins containing the same number of objects (except the last one). 
In each bin, an object is classified as photometrically variable if its 
magnitudes in the three catalogs verify the four following formulae 
at the same time:

\begin{equation}
\mid~m_1-m_2\mid+\mid~m_1-m_3\mid+\mid~m_2-m_3\mid\,\geqslant3\times\Delta~m_{tot}\label{totmag}
\end{equation}
\begin{equation}
\mid~m_1-m_2\mid\,\geqslant\Delta~m_{1,2}\label{mag1}
\end{equation}
\begin{equation}
\mid~m_1-m_3\mid\,\geqslant\Delta~m_{1,3}\label{mag2}
\end{equation}
\begin{equation}
\mid~m_2-m_2\mid\,\geqslant\Delta~m_{2,3}\label{mag3}
\end{equation}
where $\Delta~m_{tot}$ is the median of the sum of the absolute value 
of the differences of magnitude for all matched objects in the bin, 
and $\Delta~m_{i,j}$ is the median of the absolute value of the 
difference of magnitude between matched objects of catalog i and j.

Equation (\ref{totmag}) selects globally variable objects, and we ensure 
that objects are variable between each pair of catalogs with equations 
(\ref{mag1}), (\ref{mag2}) and (\ref{mag3}). This choice is more sensitive to monotically variable objects.

Photometrically variable objects whose position is closer than 10 pixels 
to a CCD defect are removed from the list, as well at those above 
the $\mu_{max}$ completeness and flagged objects. Moving objects 
are detected among single objects using a simple pipeline which 
extracts single objects with a motion compatible between image 1 and 2, 
and image 2 and 3 in a chronological order. These objects are classified as asteroids.

Finally, a reference catalog containing the classification of all objects 
in the comparison is created. This reference catalog allows the detection 
of vanished or new objects in comparisons with images taken on other nights.

\begin{figure*}[ht]
\begin{center}
\includegraphics[width=0.7\linewidth,angle=0]{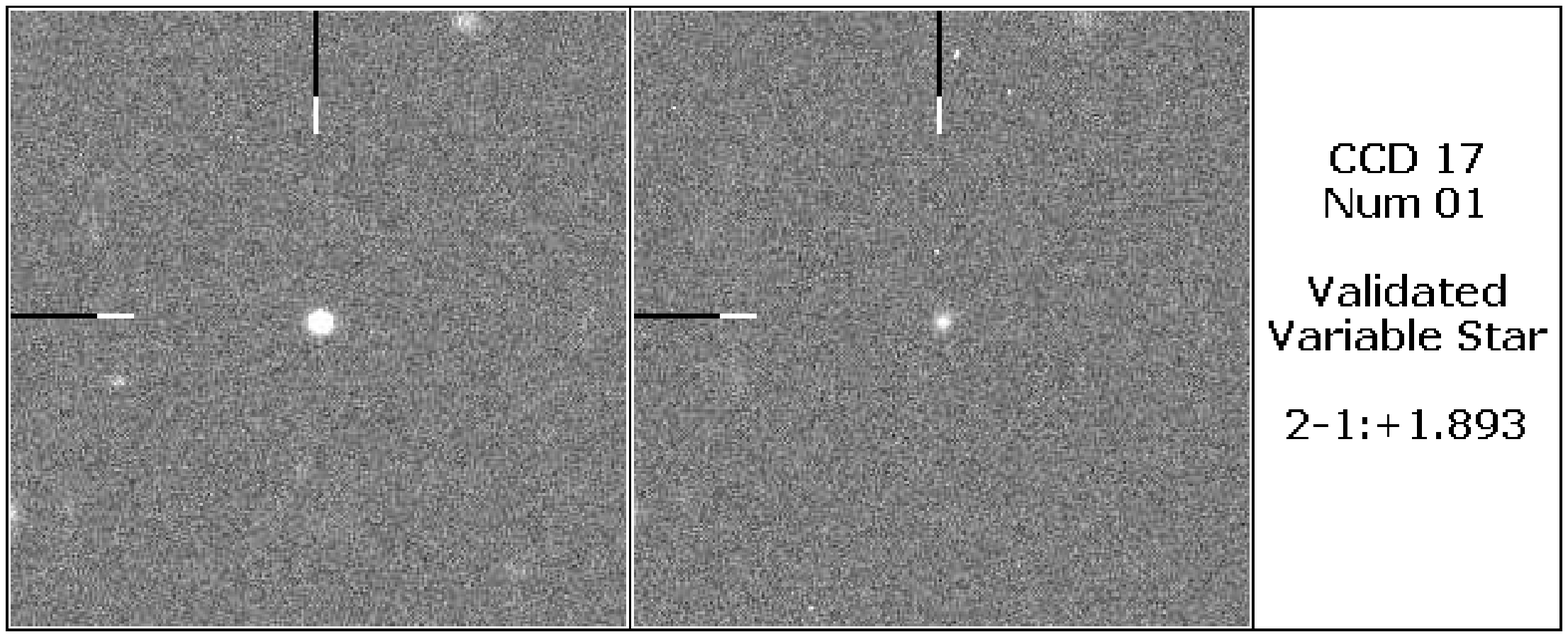}\\
\includegraphics[width=1\linewidth,angle=0]{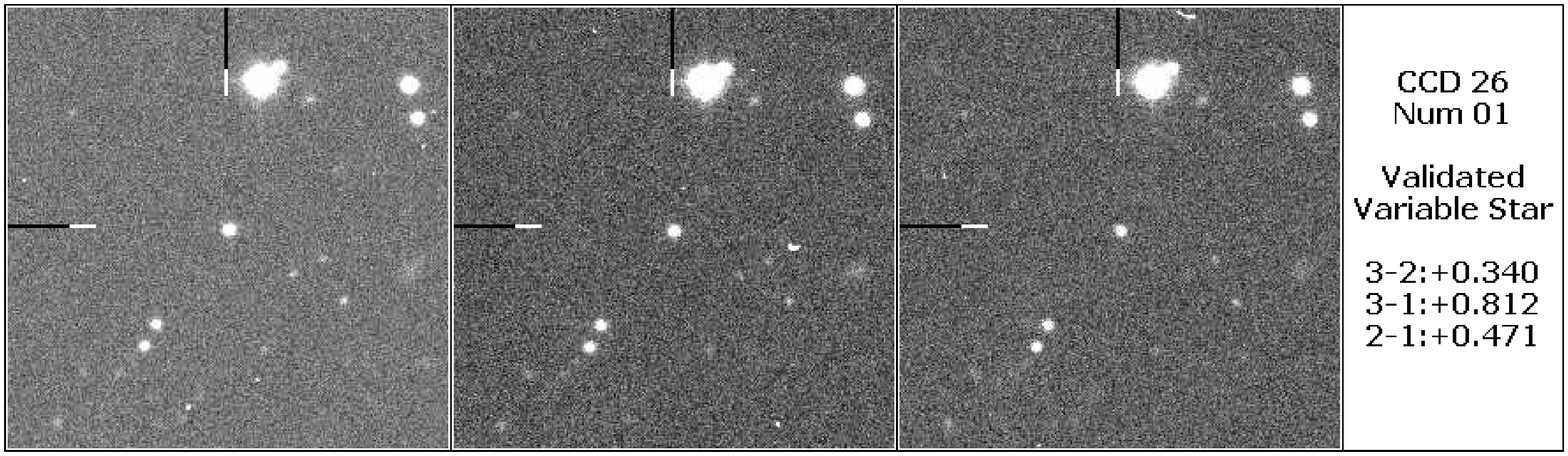}
\end{center}
\caption{These two snapshots show two characterized variable stars in 
double (top) and triple (bottom) comparisons. For each images in which 
the object is detected, a thumbnail around its position is cut, and the difference of magnitude is computed. The coordinates of the first variable star are RA = 09:01:12.39, DEC = +17:29:09.22 and its magnitude in the first image is 20.21. The coordinates of the second variable star are RA = 04:50:40.66, DEC = +21:58:21.28 and its magnitude in the second image is 21.02.}
\label{doutri}
\end{figure*}

\subsection{Double Comparisons}
Double comparisons are slightly different. One major difference is that 
the double comparisons do not allow fast and easy identification of the 
asteroids. As a consequence, these comparisons do not produce a list of astrometrically variable objects.

Like in triple comparisons, the first catalog is astrometrically matched 
to the second one, and a position tolerance is computed using pairs of nearest 
objects. Its value is similar to the one used in triple comparisons. 
Objects are classified as matched if their position difference is lower 
than the position tolerance, otherwise they are flagged as single objects. 
The  magnitude, $\mu_{max}$ and FWHM of objects of the catalog 1 are calibrated 
to those in catalog 2, using matched stars. Photometrically variable objects 
are extracted with the same method as in triple comparisons, except that 
the value compared is the absolute value on the difference of magnitude, 
which must be greater than to 4 times the median value. 
Then, we apply a correction of the difference of 
magnitude to compensate for the difference of FWHM. This correction is essential 
because many objects classified as photometrically variable are in fact 
due to seeing difference between the 2 images, as the different 
background computed in SExtractor leads to variations of the magnitude. 
On all these objects we apply cuts on magnitude, $\mu_{max}$, 
CCD defects and flag.

When one of the two catalogs is also included in a triple comparison, 
a pipeline extracts objects that have been classified as matched 
in the triple comparison, and that are classified as single 
in the double one. This procedure allows to find objects which are classified as matched in the triple images but absent in the single image. The opposite cannot be done because the single image contains asteroids which appear as new objects and cannot be rejected.

\subsection{Comparisons Output}
The results of the classification of objects in comparisons are stored 
in a database and are easily available. This comparison database 
allows the detection of disappearing or appearing objects between 
nights and runs. To summarize a comparison, variable objects are 
gathered in two HTML web pages, for photometrically and astrometrically variable objects (see Fig.~\ref{doutri}). 
These pages include a few graphics allowing an estimation of the 
quality of the comparison, a window of $250\times250$ 
pixels centered on the objects showing them in each image, 
as well as an automatic cut-out of the field in the Digital Sky Survey, 
to confirm or not the presence of bright variable objects. 
An interactive script allows a member of the collaboration 
to characterize the nature of each variable object by choosing 
between one of these categories:
\begin{enumerate}
\item No comment (A)
\item Cosmic-ray (R)
\item CCD defect (R)
\item CCD edge (R)
\item Seeing (R)
\item Contaminated\footnote{An object close to a bright object} (R)
\item Faint (R)
\item Other (R/V)
\item Galaxy (V)
\item Variable star (V)
\item Trans-Neptunian Objects (TNO) (V)
\item Candidate (V)
\end{enumerate}
Depending on this choice, the object will be classified as 
an asteroid (A), rejected (R) or validated (V), and displayed on the corresponding page. This procedure allows us to not only search GRB optical afterglows, but also to build catalogs of asteroids and photometrically variable sources. The results of this process, 
and the astrophysic characterization of the variable objects detected 
will be discussed in a forthcoming paper.

\section{Statistics}
This section presents detailed statistics for a sample of images 
taken during nearly one full year of observations. 
Statistics on catalogs represent the quality of our images, 
whereas statistics on comparisons show our efficiency to detect 
variable objects among astrophysical objects.

\subsection{Catalogs}

\begin{table}[ht]
\caption{Statistics of the \textbf{catalog creation} process.
The 9 columns give respectively the code of the observing period or run (see footnote 6), the filter f, 
the galactic latitude b,
the number of images $N_{obs}$, 
the number of square degrees observed $S_{obs}$, 
the median precision of the astrometric calibration $\delta_{pos}$,
the limiting magnitude $M_{lim}$ (see text), 
and the number of astrophysical objects found par square degree $N_{obj}/deg^{2}$.}
\label{sinstat}
\centering
\begin{tabular}{cccccccc}
\hline\hline
Run & f & b & $N_{obs}$ & $S_{obs}$ & $\delta_{pos}$ & $M_{lim}$ & $N_{obj}/deg^{2}$ \\
 & & [deg] & & [$deg^{2}$] & [''] & & [$deg^{-2}$]\\
\hline
05AQ01 & g' & +38 & 103 & 91.27 & 0.59 & 23.3 & 29\,450\\
05AQ03 & i' & +34 & 132 & 118.72 & 0.56 & 22.3 & 38\,478\\
05AQ04 & g' & +20 & 64 & 57.65 & 0.43 & 23.0 & 67\,435\\
05AQ04 & r' & +37 & 219 & 197.53 & 0.54 & 22.6 & 43\,278\\
05AQ05 & g' & +22 & 86 & 77.18 & 0.45 & 22.8 & 50\,314\\
05AQ05 & r' & -16 & 13 & 11.68 & 0.56 & 22.0 & 225\,669\\
05AQ05 & i' & -47 & 30 & 27.07 & 0.60 & 22.3 & 41\,937\\
05BQ11 & g' & -12 & 50 & 45.10 & 0.45 & 23.6 & 49\,156\\
05BQ13 & r' & +32 & 83 & 74.72 & 0.54 & 22.8 & 37\,832\\
05BQ13 & i' & -16 & 52 & 46.92 & 0.46 & 22.4 & 50\,082\\
06AQ01 & r' & +39 & 74 & 66.73 & 0.59 & 22.3 & 30\,095\\
06AQ01 & i' & -12 & 52 & 46.92 & 0.44 & 22.8 & 76\,439\\
\bf All & \bf- & \bf- & \bf958 & \bf861.49 & \bf0.52 & \bf22.7 & \bf46\,821\\
\hline
\end{tabular}
\end{table}

\begin{figure}[ht]
\centering
\resizebox{\hsize}{!}{\includegraphics{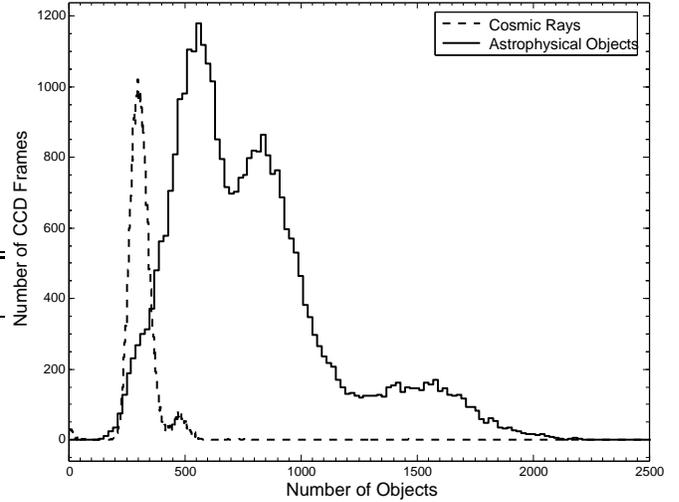}}
\caption{Histogram of the total number of astrophysical objects (solid line), 
and of the cosmic-ray hits (dashed line) in all the CCD frames studied.
 The nearly constant number of cosmic-ray hits, about 300 per CCD frame, is explained by the quasi constancy
of the exposure time. The number of astrophysical objects, on the other hand
appears much more variable as it depends strongly on the filter, and on 
the galactic latitude of the observations.
}
\label{usnomag}
\end{figure}

Table~\ref{sinstat} presents some catalog statistics based on 
958 observations. If we consider that the field of view 
of MegaCAM is $0.96\deg\times0.94\deg$, the total sky 
coverage $S_{obs}$ is 864.5\,deg$^{2}$. 
By properly processing 861.5\,deg$^{2}$, the RTAS has an efficiency 
of 99.65\% for the catalog creation process. 
We note that the USNO-A2.0 matching precision $\delta_{pos}$ 
is always better than 0.6\arcsec. The completness magnitude $M_{lim}$, which is 
strongly dependent on the filter used, is roughly distributed 
from $r'=22$ to $g'=23.6$, with a median value of 22.7. 
With the exception of the r' filter images of 
05AQ05\footnote{A run at the CFHT is named according to the following 
designation: two numbers for the year, two letters for the semester 
(AQ for the first semester and BQ for the second one), and two 
numbers for the period of observation} which were pointing near 
the galactic center, the total number of objects per 
square degree $N_{obj}/deg^{2}$ is about 50\,000, depending 
on the filter, the seeing, and the observed region of the sky. 
Fig~\ref{usnomag} summarizes the efficiency of the classification part 
of the catalog creation process (see Fig~\ref{sort}). 
It can be noticed that, while the number of astrophysical sources strongly varies, the number of 
cosmic-ray hits per CCD frame is nearly constant, except for 05BQ11, which has an exposure time twice than usual.

\subsection{Triple Comparisons}
\begin{figure*}[ht]
\begin{center}
\includegraphics[width=0.495\linewidth,angle=0]{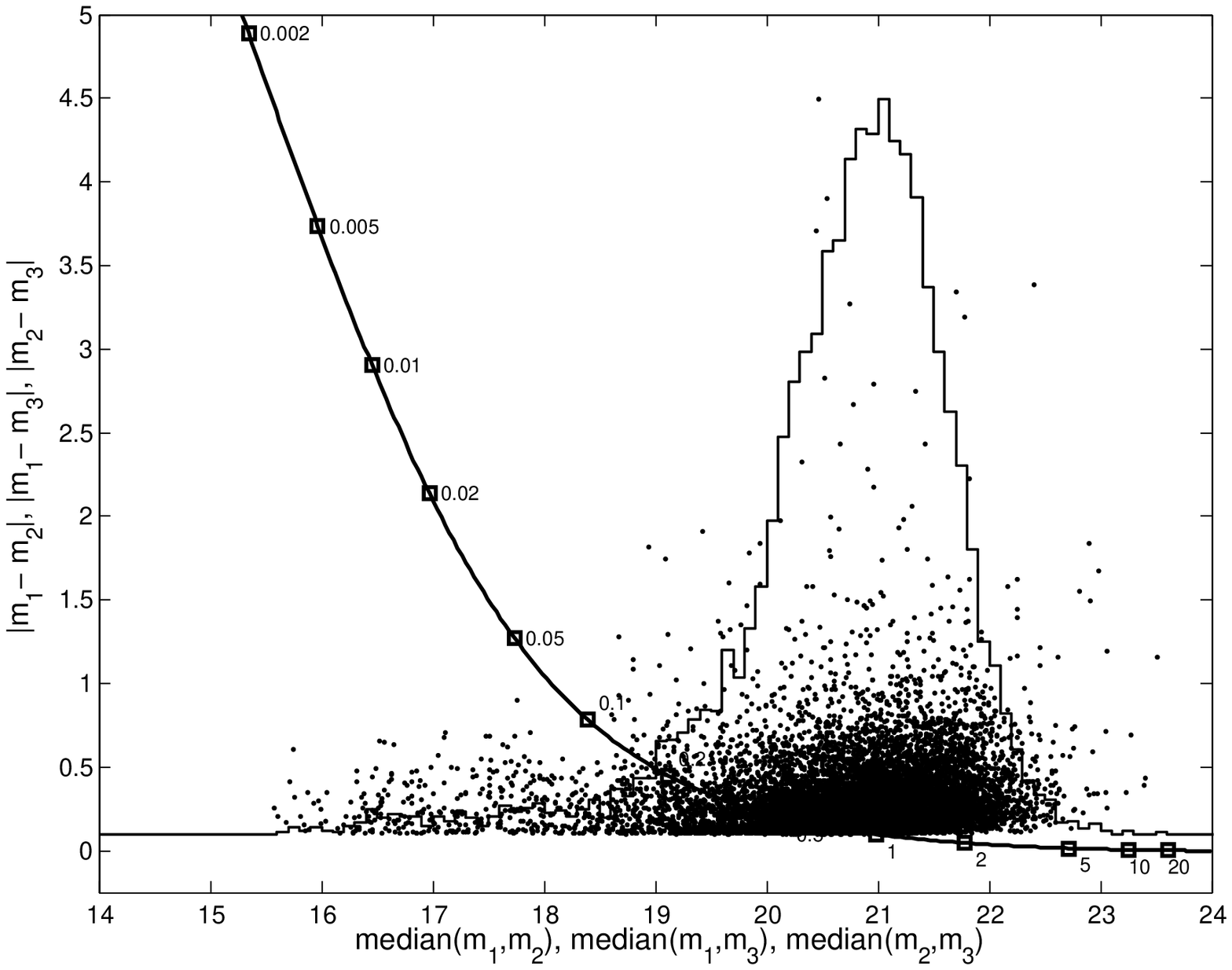}
\includegraphics[width=0.495\linewidth,angle=0]{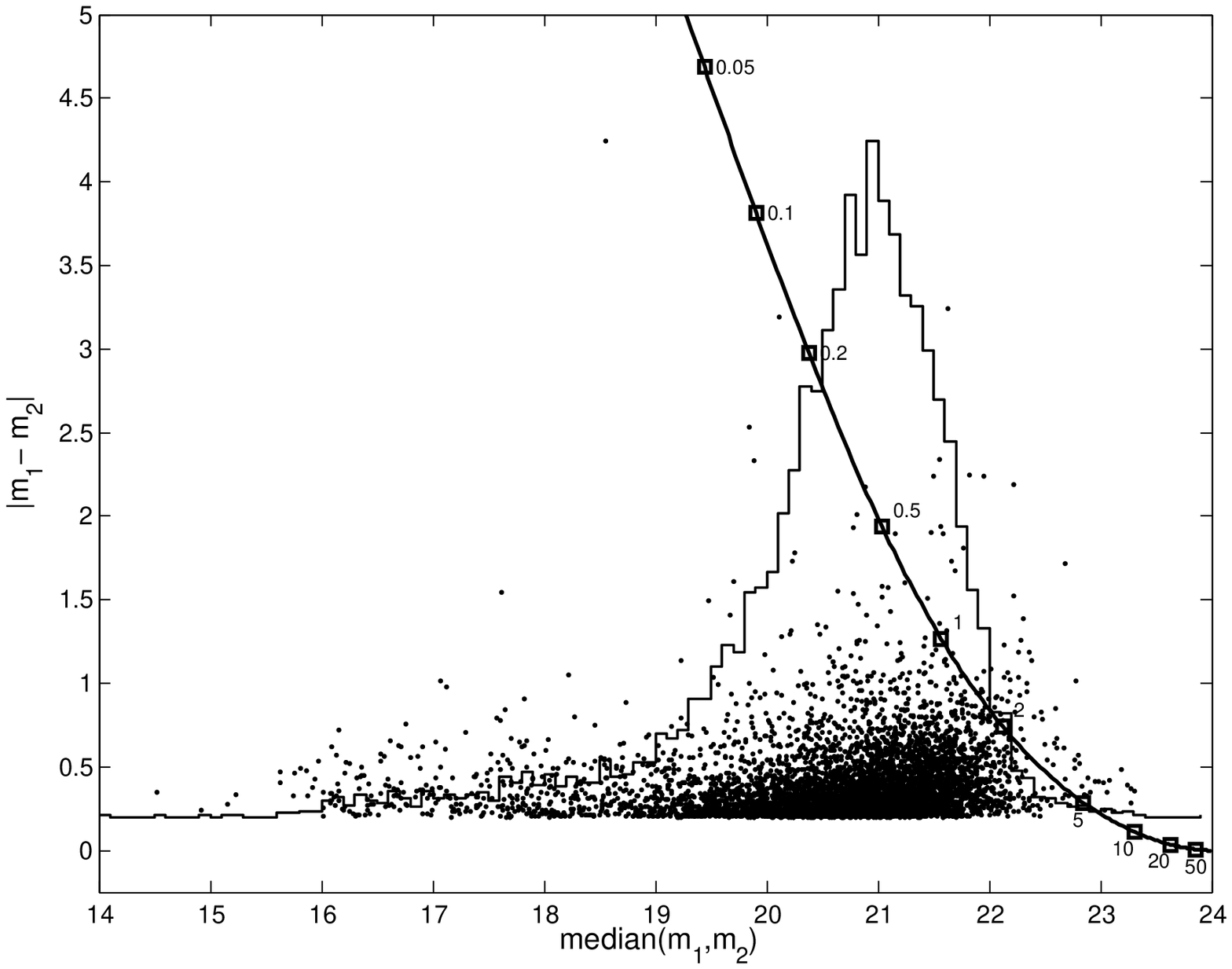}
\end{center}
   \caption{Objects detected variable by our automatic software, in triple 
(left) and double (right) comparisons.
Each point represents one object in the comparison of a pair of images.
The x-axis shows the median magnitude while the y-axis shows the
magnitude difference between the two images. 
In the triple comparisons each object is represented by three
points corresponding to the three possible pairs of images.
The distribution of these variable objects is shown to illustrate 
the domain of sensitivity of our search (in magnitude and $\Delta mag$).
For comparison we have also shown the track of a typical afterglow 
($\alpha=1.2$, $M_{1}=21$ and $M_{host}=24$, see Sect.7), 
as a function of its age (in days) at the time of the first
observation.}
   \label{tridouvar}
\end{figure*}

As there is only a small chance of detecting GRB afterglows 
in triple comparisons (see Fig~\ref{tridouvar}), they are mainly 
used to determine the true nature of objects, and to create 
the reference catalogs for the double comparisons. This is important because we expect a lot of asteroids and few variable objects, since 
the Very Wide Survey points to ecliptic plane and 
images are taken only 1 hour apart.

\begin{table}[ht]
\caption{Statistics for the \textbf{triple comparisons} process.
The 8 columns give respectively the code of the observing period, 
the filter, 
the number of comparisons $N_{tri}$, 
the sky area involved in the comparisons $S_{tri}$, 
the limiting magnitude of the comparisons $M_{lim}$, 
the number of asteroids found in the comparisons  $N_{ast}$,
the number of variable sources found in the comparisons $N_{var}$,
and the number of variable candidates per 10$^6$ astrophysical 
objects found by the program $N_{6}$.}
\label{tristat}
\centering
\begin{tabular}{c c c c c c c c}
\hline\hline
Run & Flt & $N_{tri}$ & $S_{tri}$ & $M_{lim}$ & $N_{ast}$ & $N_{var}$ & $N_{6}$\\
 & & & [$deg^{2}$] & & & & \\
\hline
05AQ01 & g' & 24 & 20.83 & 23.1 & 1\,072 & 242 & 744\\
05AQ03 & i' & 31 & 27.55 & 22.0 & 912 & 362 & 585\\
05AQ04 & g' & 16 & 14.36 & 22.9 & 618 & 398 & 537\\
05AQ04 & r' & 50 & 44.94 & 22.5 & 2\,543 & 842 & 657\\
05AQ05 & g' & 16 & 14.31 & 22.6 & 488 & 302 & 599\\
05AQ05 & i' & 10 & 9.02 & 22.0 & 288 & 88 & 387\\
05BQ11 & g' & 7 & 6.32 & 23.4 & 274 & 91 & 624\\
05BQ13 & r' & 20 & 17.95 & 22.6 & 1\,129 & 270 & 746\\
05BQ13 & i' & 13 & 11.73 & 22.1 & 507 & 220 & 505\\
06AQ01 & r' & 18 & 16.19 & 22.1 & 749 & 152 & 638\\
06AQ01 & i' & 13 & 11.73 & 22.7 & 800 & 542 & 762\\
\bf All & \bf- & \bf218 & \bf194.94 & \bf22.5 & \bf9\,380 & \bf3\,509 & \bf628\\
\hline
\end{tabular}
\end{table}

As shown in Table~\ref{tristat}, 194.94 square degrees were 
compared in 218 triple comparisons. The magnitude completeness 
$M_{lim}$ is brighter than in the catalogs, because only matched 
objects were taken into account, so it represents in fact the 
completeness magnitude of the worst image of the triple comparison. 
9\,380 asteroids have been detected amongst 7\,910\,214 single 
objects\footnote{Since a single object appears in only one image, 
this value, which includes cosmic rays, has to be divided by 3 
to be compared to the matched objects number}. 3\,509 matched 
objects were classified as variable amongst $5.10^6$ matched objects (only 1 per 1593 objects). 
In order to compare this value for all runs and filters 
independently from the number of objects, we used $N_{6}$, which is the 
number of variable objects per 1\,000\,000 matched objects. 
For the triple comparisons, $N_{6}$ is usually around 400 and 700. 
The difference of magnitude of variable objects as a function 
of their magnitude can be seen in Fig~\ref{tridouvar}. 
90\% of variable objects are detected with a variation of less 
than 0.57 magnitude and 99\% have a variation below 1.21. 
Objects with strong magnitude variation are usually variable 
stars or asteroids superimposed on a faint object.

\subsection{Double Comparisons}
\begin{table}[ht]
\caption{Statistics for the \textbf{double comparisons} process.
The columns are identical to those of Table \ref{tristat}, except for 
$N_{sin}$ which is the number of new or vanished objects. This number is higher for the g' filter images of 05AQ04 and 05AQ05 because halos of bright stars generate more fake objects in this filter.}
\label{doustat}
\centering
\begin{tabular}{c c c c c c c c}
\hline\hline
Run & Flt & $N_{dou}$ & $S_{dou}$ & $M_{lim}$ & $N_{sin}$ & $N_{var}$ & $N_{6}$\\
 & & & [$deg^{2}$] & & & & \\
\hline
05AQ01 & g' & 28 & 24.84 & 22.6 & 10 & 317 & 1281\\
05AQ03 & i' & 24 & 21.66 & 22.3 & 2 & 463 & 790\\
05AQ04 & g' & 16 & 14.41 & 23.0 & 258 & 564 & 847\\
05AQ04 & r' & 62 & 55.87 & 22.5 & 132 & 1\,231 & 796\\
05AQ05 & g' & 34 & 30.36 & 22.8 & 296 & 881 & 807\\
05AQ05 & r' & 3 & 2.66 & 21.5 & 0 & 80 & 132\\
05AQ05 & i' & 5 & 4.51 & 22.2 & 5 & 68 & 551\\
05BQ11 & g' & 7 & 6.32 & 23.6 & 15 & 94 & 522\\
05BQ13 & r' & 20 & 17.93 & 22.7 & 34 & 568 & 1387\\
05BQ13 & i' & 15 & 13.54 & 22.2 & 62 & 329 & 720\\
06AQ01 & r' & 20 & 17.95 & 22.5 & 64 & 449 & 1387\\
06AQ01 & i' & 13 & 11.68 & 22.4 & 97 & 656 & 980\\
\bf All & \bf- & \bf247 & \bf221.71 & \bf22.5 & \bf975 & \bf5\,700 & \bf825\\
\hline
\end{tabular}
\end{table}

247 double comparisons were performed for a total of 221.71 
square degrees. The completeness magnitude $M_{lim}$ is a little 
better than the one in the triple comparisons statistics, because 
the best image of the triplet is selected. Except for comparisons 
of the g' filter of 05AQ04 and 05AQ05, the number of objects 
detected as vanished or new, $N_{sin}$, is only a few per comparison, 
amongst a total of 7\,638\,326 single objects. $N_{6}$, the number 
of variable objects per $10^6$ matched objects, is higher 
than in triple comparisons, reaching a mean of 825. 
This is mainly due to two factors: first, inter-night comparisons 
detect objects which are variable on timescales of hours to days, 
including the objects detected in intra-night comparisons, and second, images taken 2 days 
apart are more sensitive to strong variations of the climatic conditions, 
resulting in more fake detections of variable objects due to seeing differences. 
Nevertheless, the number of detected variable objects stays low, 
with only 1 variable object for 1212 objects.

\subsection{Characterization}
\begin{table}[ht]
\caption{Characterization of variable objects of 05AQ01}
\label{caracstat}
\centering
\begin{tabular}{| l | c c | c c |}
\hline
& \multicolumn{2} {c |}{Triple comparisons} & \multicolumn{2} {| c |}{Double comparisons} \\
 & Number & \%$^{a}$ & Number & \%$^{a}$\\
\hline
No comment & 1 & 0.41 & 0 & 0\\
Cosmic-ray & 10 & 4.13 & 31 & 9.48\\
CCD defect & 42 & 17.36 & 100 & 30.58\\
CCD edge & 2 & 0.83 & 9 & 2.75\\
Seeing & 113 & 46.69 & 129 & 39.45\\
Contaminated & 55 & 22.73 & 9 & 2.75\\
Faint & 0 & 0 & 5 & 1.53\\
Other & 0 & 0 & 0 & 0\\
Galaxy & 2 & 0.83 & 5 & 1.53\\
Variable star & 17 & 7.03 & 39 & 11.93\\
TNO & 0 & 0 & 0 & 0\\
Candidate & 0 & 0 & 0 & 0\\
\bf Rejected & \bf222 & \bf91.7 & \bf283 & \bf86.54\\
\bf Validated & \bf19 & \bf7.8 & \bf44 & \bf13.46\\
\hline
\end{tabular}
\begin{list}{}{}
\item[$^{a}$] percent of all variable objects
\end{list}
\end{table}
In Table~\ref{caracstat} we provide an example of the characterization 
by a member of the collaboration of objects selected as variable by 
the automatic process for the run 05AQ01. About 90\% of the variable objects 
are false detections. Most of them are due to seeing problems, 
CCD defects or contaminated objects. The validated objects are 
mostly variable stars. During this run we have not found any 
trans-neptunian object, and no object was interesting enough 
to be characterized as an afterglow candidate.

These statistics clearly show our conservative point of view about 
the selection of variable objects. The number of false detections 
like CCD defects or seeing problems could easily be reduced, 
at the expense of a lower sensitivity for afterglow detection. If there is little chance to detect a GRB afterglow within triple comparisons, 
their detection is fully possible in double comparisons. 
As shown in Fig~\ref{tridouvar}, a typical afterglow would be 
detected as a variable object with $m\sim21.5$, $\Delta~m\sim1$ 
and $\Delta~t\sim1$ day in the double comparison.

\section{Estimations and Comparisons with other surveys}
In this section, we compare the performance of our search for orphan 
afterglows using the Very Wide Survey with previous attempts. 
We do not discuss the estimation of the collimation factor, 
because all images taken have not been analyzed yet. 
This estimation will be the main purpose of a second paper.

We evaluate the performance of each survey along the following lines: 
we draw a sample of GRB afterglows and compute the number of them
that will be detected by each survey taking into account 
its depth and strategy of observation.

\begin{figure}[ht]
\centering
\includegraphics[width=0.495\linewidth,angle=0]{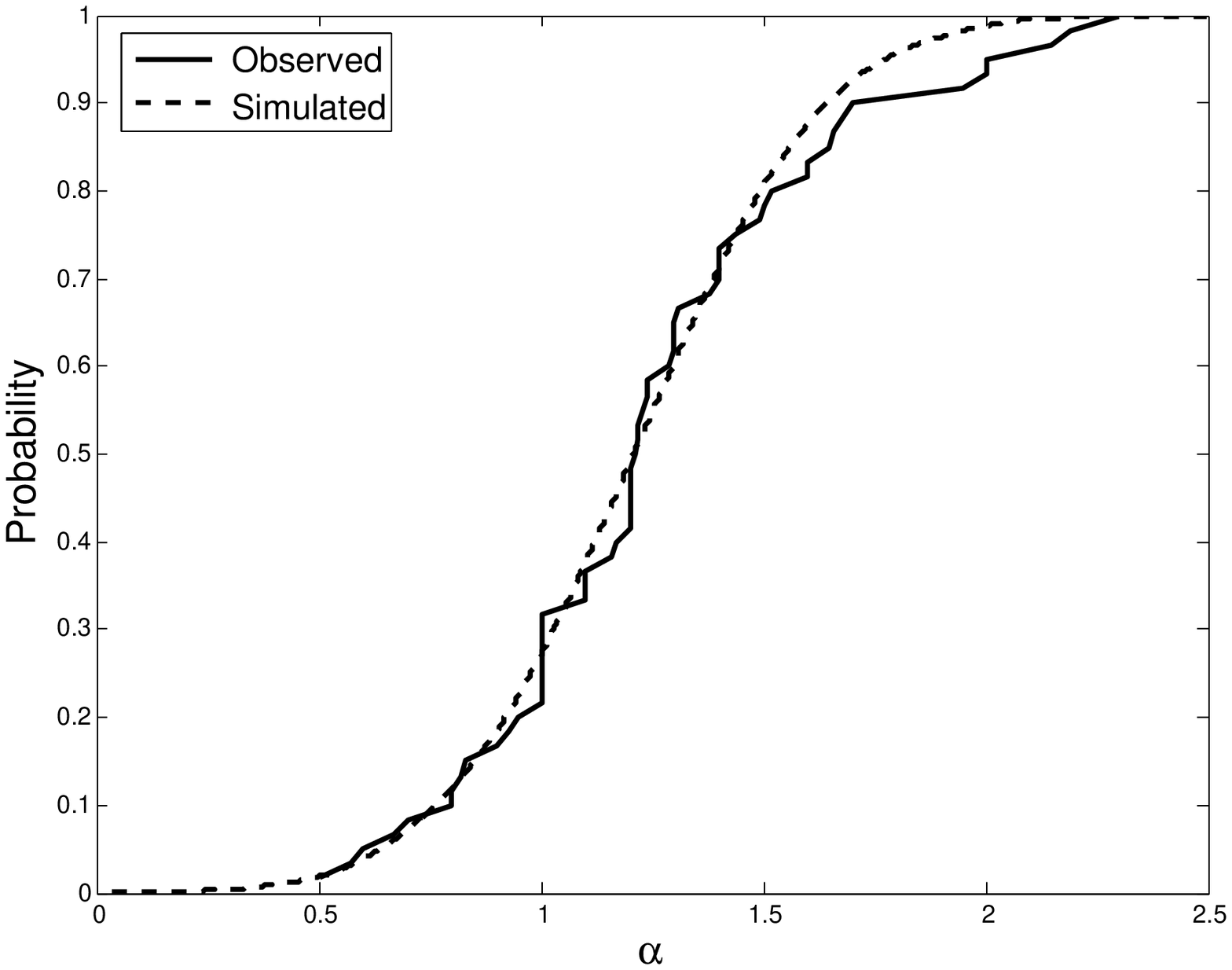}
\includegraphics[width=0.495\linewidth,angle=0]{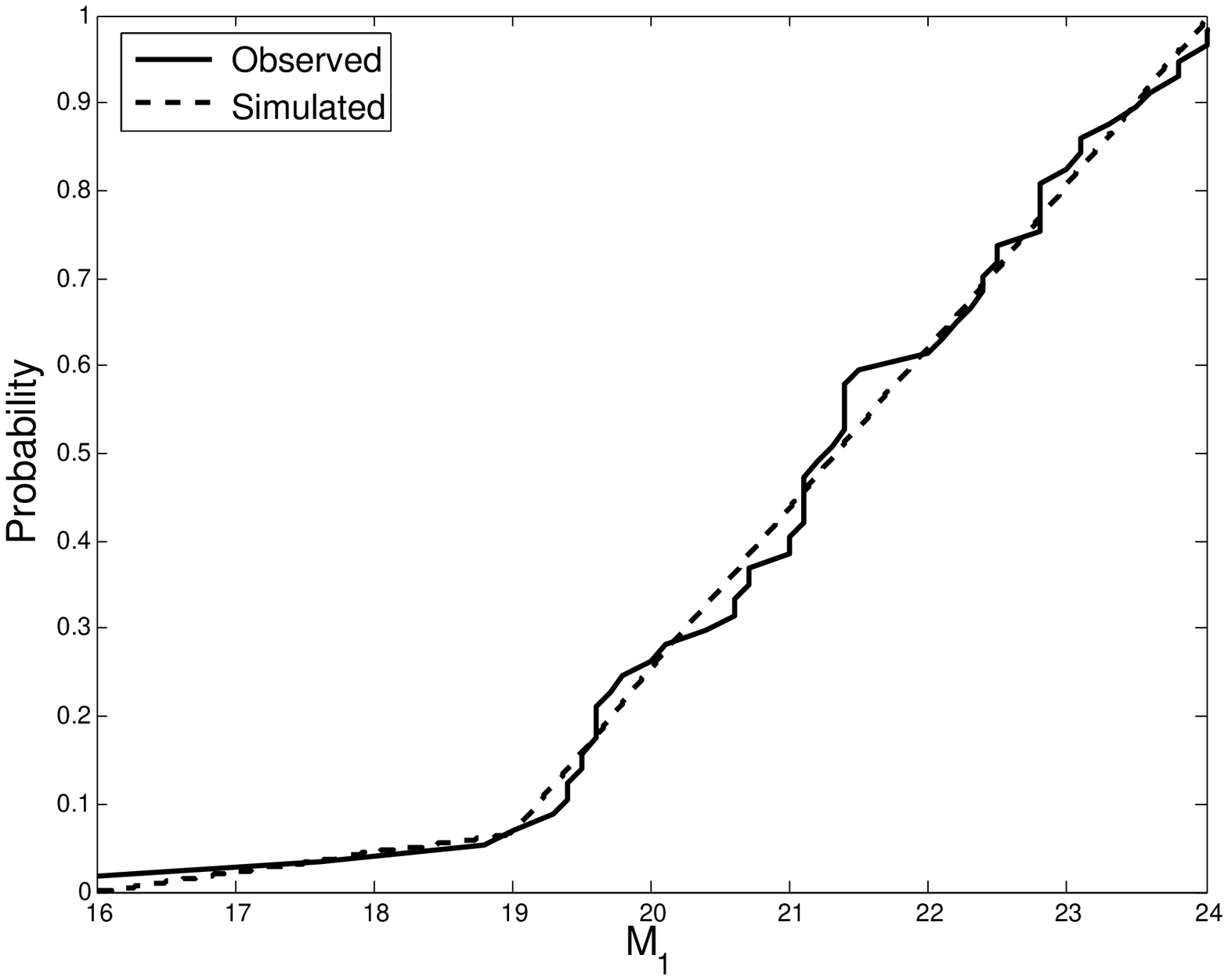}
   \caption{These two figures show the normalized distributions 
of $\alpha$ (left) and $M_{1}$ (right) for 60 observed afterglows 
and the distribution we choose to fit the data. Since there is no 
correlation between $\alpha$ and $M_{1}$, these values can be 
drawn independently}
   \label{alpmag}
\end{figure}

We simulated afterglows with 5 random parameters: burst date, 
right ascension and declinaison, temporal decay slope $\alpha$ 
and magnitude at one day $M_{1}$. Light curves of simulated afterglows 
were chosen to be simple power law functions. The two intrinsic parameters 
$\alpha$ and $M_{1}$ were randomly drawn using probability laws 
fitted on 60 observed afterglows taken from various GCN notices 
(see Fig.~\ref{alpmag}). GRBs are produced at random times and 
isotropically on the sky.

The number of afterglows generated depends on 2 parameters: 
$N_{\gamma}$ the number of bursts whose jet is directed towards 
the earth, which is independent of the observational strategy, 
and $\beta$ the collimation factor, which is simply the total 
number of bursts divided by $N_{\gamma}$. $\beta$ strongly 
depends on the limiting magnitude of the observations 
(Totani \& Panaitescu \cite {totani02}). So, the number of 
afterglows generated for each survey is simply defined by 
$N_{\gamma}\times\beta$, where we choose $N_{\gamma}=800$, 
following Rau et al. (\cite {rau}).

Each survey is described by the following parameters (see table~\ref{compstrat}):
\begin{itemize}
\item[-] $S_{obs}$, the mean sky coverage in deg$^{2}$. 
In the computation of $S_{obs}$, we add up the areas of the images 
of the same field which are separated in time by more than the mean time of visibility 
of the afterglows.
\item[-] $\delta t$, the time between the two observations of each pair of images.
\item[-] $M_{lim}$, the mean completeness magnitude of observations
\end{itemize}

By using the completeness magnitude of each survey and the light-curves of simulated afterglows, we are able to derive their mean time of visibility $t_{vis}$, which is the time below which 50\% of the afterglows remain visible in the images of the survey.

\begin{table}[ht]
\caption{Comparison of 4 programs dedicated to the search of GRB optical afterglows, $\beta$ have been chosen 
according to Totani \& Panaitescu (\cite {totani02})}
\label{compstrat}
\centering
\begin{tabular}{c c c c c c c}
\hline\hline
Survey & $\beta$ & $S_{obs}$ & $t_{vis}$$^{a}$ & $\delta t$ & $M_{lim}$ & $N_{exp}$$^{b}$ \\
 & & [deg$^{2}$] & [days] & & &\\
\hline
 ROTSE-III$^{c}$ & 2 & 65\,550 & 0.07 & 0.02 days & 18 & 0.6 \\
 Rau et al.$^{d}$ & 15 & 55$^{e}$ & 3.5 & 3 days & 23 & 0.3 \\
 Very Wide & 11 & 1\,178 & 2.5 & 2 days & 22.5 & 4.6 \\
 Optimal & 21 & 250 & 7.5 & 7 days & 24 & 5.6 \\
\hline
\end{tabular}
\begin{list}{}{}
\item[$^{a}$] Mean time of visibilty of afterglows
\item[$^{b}$] Number of afterglows expected
\item[$^{c}$] Rykoff et al. (\cite {rykoff05})
\item[$^{d}$] Rau et al. (\cite {rau})
\item[$^{e}$] $\sum_{i=1}^7~\big[(N_{sub, i}\times\frac{12}{\sum_i N_{sub, i}})\times(\frac{N_{night, i}}{3.5})\big]$, where $N_{sub, i}$ is the number of subfield and $N_{night, i}$ the number of nights of observation of each field
\end{list}
\end{table}

We choose to compare the observational strategy of the Very Wide Survey 
with 2 other surveys specially dedicated to GRB optical afterglow detection.

The survey used by Rykoff et al. (\cite {rykoff05}) has been 
performed with the ROTSE-III telescope. It has an extra wide sky 
coverage, but low sensitivity; that's why the collimation factor 
is modest. Each field is observed twice within 30 minutes, and is 
considered as independent. This strategy is optimized for early 
afterglow detection. As shown in Table~\ref{compstrat}, half of the 
afterglows become undetectable about 2 hours after the burst. 
During this survey, 23000 sets were observed with a mean field 
size of 2.85 deg$^{2}$, so $S_{obs}=65500$. The distribution of 
magnitude along all sets in Rykoff et al. (\cite {rykoff05}) gives 
a completness magnitude $M_{lim}\sim18$. 
According to Totani \& Panaitescu (\cite {totani02}), $\beta$ is 
equal to 2 at this magnitude. By launching the simulation 50 times 
using this observing strategy, we can estimate the number of afterglows 
expected to be about 0.6. This is consistent with their analysis, 
since no GRB afterglow candidate has been discovered.

Another attempt has been performed by Rau et al. (\cite {rau}) with 
the WFI Camera at the 2.2m MPI/ESO telescope in La Silla. 
The observational strategy was based on 7 fields, divided in sub-fields, 
and with multiple observations during a maximum of 25 nights in a row. 
Although 12 deg$^{2}$ were really observed, we upgrade the sky coverage 
to 55 independent square degrees observed twice within 3 days, 
using the mean time of visibility of afterglows, which is 3.5 days 
(see Table~\ref{compstrat}), and considering the difference of the observational 
strategy for each field (see Table~\ref{compstrat}$^{e}$). 
We choose to use the completeness magnitude given by Rau et al., 
which is $r=23$, although it seems to be in fact the limiting magnitude. 
This magnitude gives $\beta=15$ in Totani \& Panaitescu (\cite {totani02}). 
The simulation was launched 50 times with these values. We estimate 
the number of afterglows expected in this survey to be 0.3 according 
to our simulation. This value is comparable with the results of their 
study: one object similar to an afterglow was found, although it has been
later confirmed as a flare star (Kulkarni \& Rau \cite {kulkarni06}).

Since the beginning of the Very Wide Survey, 4632 images were 
taken on 612 different fields of the sky. 1178 independent fields 
of $0.96\times0.94\,deg^{2}$ have been observed at a mean magnitude 
of $r=22.5$, at which $\beta=15$ according to Totani \& Panaitescu 
(\cite {totani02}). By using these values in 50 simulations, 
we estimate the number of afterglows in all of these images to be 4 to 5.

This simple simulation shows that the Very Wide Survey is to date the 
most adapted survey for the search of optical afterglows. 
Based on the predictions of Totani \& Panaitescu (\cite {totani02}),
we expect about 4 afterglows in the entire survey, ten times more 
than the survey of Rau et al..

\section{Discussion of an optimal observational strategy}
Although the observational strategy of the Very Wide Survey has not been built to search for GRB optical afterglows, our simulations show that the number of afterglows expected in this survey is ten times higher than in other dedicated programs. In this section, we will take advantage of the experience acquired from this work to further discuss the optimal observational strategy.

Given the rarity of optical GRB afterglows, the choice of an observational strategy is crucial to optimize their detection. Each strategy can be divided in two distincts part: the spatial part which defines the size and depth of the area observed, and the temporal part which defines the number of observations for each field and the time between observations. Since afterglows are very rare objects, and since their light-curves decrease like a power-law with time, a compromise has to be found between the depth and the width of the survey. A wide shallow survey favours the detection of "early" and bright afterglows, but our simulation based on the ROTSE-III survey clearly shows that the chance of detecting such objects is very low because the beaming factor remains high during the early time of the burst. On the contrary, a deep survey favours late and faint detections. But, while the afterglows are much more numerous at faint magnitude\footnote{The power law decay implies that the beaming factor is low and the afterglows are faint during 90\% of their lifetime.}, their detection is made more difficult by their slow decay and by the presence of the host galaxy. However, in our simulation, this kind of survey seems to be the most appropriate to search for optical afterglow.

Concretely, parameters that have to be defined to build an optimal survey are the sky coverage, the global observing strategy (number of observations and delay between them), and the depth of the observations. As we mentioned in the previous paragraph, when an afterglow reaches a certain magnitude, it starts to be hidden by its host galaxy, and its magnitude variation is not detectable anymore; therefore the mean magnitude of observed hosts of afterglows, $r=24$, seems to be a good value for the completeness magnitude. Due to the power-law decrease of its light curve, a typical afterglow doesn't have strong magnitude variations at high magnitudes. So the observations have to be sufficiently spaced in time in order to have a difference of magnitude that allows the detection of the variability of the afterglow. At $r=24$, the mean time of visibility of afterglows is about 7.5 days (see Table~\ref{compstrat}). The maximum time between the two main observations can be chosen to be 7 days, but it can also be reduced to a few days in case of climatic or priority problems without any strong inconveniences.

Our experience clearly shows the necessity of a reference catalog in order to check the presence of variable objects and to detect new ones. When possible, the observed fields must be chosen to be part of an available survey at least as deep as the completeness magnitude, otherwise the fields have to be individually observed before the main observations within the survey, so these observations can be used to construct a reference catalog. Within the main observations, it is crucial to be capable to detect new or vanished objects, because a significant number of afterglows may appear or disappear between the two main observations. The search for such objects can only be processed by using the reference catalog, but, in order to be sure that the object is a non-moving stationary astrophysical source, there should be at least two observations for each main observation, taken during the same night. This will allow to fill the gaps between CCD frames and to construct an internal reference catalog, which will be very useful to characterize selected objects. Also a good idea is to refrain from pointing the ecliptic plane in order to avoid asteroids.

Since colors help neither for the detection\footnote{Images taken with different filters cannot be compared.} nor for the characterization of the sources, all the observations can be done with the same filter. As this time of our investigation, we are not able to select a favourite filter, but since afterglows are at high redshift, a red filter would be a good choice. While such a survey is sufficient for the detection of the afterglows, a follow-up is still needed to confirm the nature of the detected objects. Since the confirmation of the variability of the object will most of the time take place during the second main observation, a fast identification is needed for confirmation with X-ray telescopes or big optical telescopes on the ground. A spectral analysis is also still conceivable, because the completeness magnitude $r=24$ corresponds to the limit magnitude below which a spectrum can be obtain with 8-10m class telescopes.

Given the above considerations, we can design an "optimal" survey that will have $M_{lim}\sim24$, $\delta t\lesssim7$\,days and $S_{obs}\sim250$\,deg$^{2}$. This survey would require about 7 full nights of observation with a CCD imager similar to MegaCAM at the CFHT, and 6 afterglows would be expected.

We conclude this section with an observation which has been a surprise for us: the very low background of astronomical
sources which vary like GRB afterglows. It is interesting to note that a series of 3 to 5 exposures with a single filter was sufficient to eliminate nearly all the events which were strongly variable, like GRB afterglows. An essential ingredient in this task is the availability of at least one image taken months before, to check the existence in the past of the variable sources detected by the software.

\section{Conclusion}
In this paper, we have presented a new untrigerred search for optical GRB afterglows within the images of the Very Wide Survey at the Canada France Hawaii Telescope. In this survey, each field is observed three times during the same night and once 1 or 2 days later, allowing the detection of variable objects. Up to now, 1\,178 independent fields of nearly 1\,deg$^{2}$ have already been observed, down to $r'=22.5$.

We have described in details the Real Time Analysis System "Optically Selected GRB Afterglows". This automatic pipeline, specially dedicated to an untriggered search of GRB afterglows, extracts variable, new and vanished objects, by comparing two or three catalogs of objects of images of the same field of the sky. Variable objects are displayed on a web page to be characterized by a human.

In order to quantify the efficiency of the process, statistics were computed on nearly one full year of observations. These statistics clearly show the quality of the images and of their processing, as well as our capability to detect variable objects within these images. Five to ten of 10\,000 objects are classified as variable by the process, but only 10\% of them are true astrophysical variable objects. In addition, we detect about 50 asteroids and a few new or vanished objects by comparison.

We finally performed a simulation of afterglow detection in order to compare our search with previous attempts, which have been unsuccessful. Our simulated  afterglows are based on 60 real afterglows described in the GCNs. According to this simulation, the Very Wide Survey has an efficiency, which is ten times higher than previous searches.

We discussed an optimal survey for the search for GRB afterglows, based on the experience acquired from this work. Many considerations were taken into account, like the observational flexibility, the detection improvement and the follow-up opportunities for the confirmation of the object. This optimal survey, which can be completed with a few nights of observations with a telescope similar to the CFHT, will allow the detection of about 6 GRB afterglows according to the predictions of Totani \& Panaitescu (\cite {totani02}). Our current experience demonstrates that the background of variable sources behaving like GRB afterglows is very low, allowing efficient searches based on the acquisition of few images of the same region of the sky taken hours to days apart with a single filter, with a reference taken 1 or 2 months before.

Since the RTAS is operational since November 2004, only one half of the fields observed within the Very Wide Survey have been searched for afterglows in real-time. Although a few objects that behaved like GRB afterglows have been found, none of them revealed to be a real afterglow. In the case that no afterglow is detected in the whole Very Wide Survey images, we will derive an upper limit of 6 orphan afterglows per 1 on-axis afterglow down to magnitude $r'=22.5$. This value is consistent with the predictions of Totani \& Panaitescu (\cite {totani02}) and Nakar et al. (\cite{nakar02}).

In a forthcoming paper, we will present the complete analysis of all variable objects found in the Very Wide Survey images. We will also discuss the estimation of the collimation factor of gamma-ray bursts.

\begin{acknowledgements}
We would like to thank everyone at the CFHT for their continuous support, especially Kanoa Withington.
We also thank the Observatoire Midi-Pyr\'en\'ees for having funded the RTAS.
\end{acknowledgements}


\begin{thebibliography}{}

\bibitem[2004]{becker}
Becker, A. C. et al. 2004, ApJ, 611, 418

\bibitem[2003]{bloom03}
Bloom, J. S., Frail, D. A., Kulkarni, S. R. 2003, ApJ, 594, 674

\bibitem[2002]{dalal02} 
Dalal, N., Griest, K., \& Pruet, J.\ 2002, \apj, 564, 209 

\bibitem[2001]{frail01}
Frail, D. A. et al. 2001, ApJ, 562, L55

\bibitem[2002]{galyam02}
Gal-Yam, A., et al. 2002, PASP, 114, 587

\bibitem[2006]{galyam06} 
Gal-Yam, A., et al.\ 2006, \apj, 639, 331 

\bibitem[1999]{greiner99} 
Greiner, J., Voges, W., Boller, T., \& Hartmann, D.\ 1999, \aaps, 138, 441 

\bibitem[2003]{groot03} 
Groot, P.~J., Vreeswijk, P. M., Huber, M. E. et al.\ 2003, \mnras, 339, 427 

\bibitem[1999]{harrison99} 
Harrison, F.~A., Bloom, J. S., Frail, D. A. et al.\ 1999, \apjl, 523, L121 

\bibitem[2002]{huang02} 
Huang, Y.~F., Dai, Z.~G., \& Lu, T.\ 2002, \mnras, 332, 735

\bibitem[2002]{kehoe02} 
Kehoe, R., Akerlof, C., Balsano, R. et al.\ 2002, \apj, 577, 845 

\bibitem[1999]{kulkarni99} 
Kulkarni, S.~R., Djorgovski, S. G., Odewahn, S. C et al.\ 1999, \nat, 398, 389 

\bibitem[2006]{kulkarni06}
Kulkarni, S.~R., \& Rau, A., 2006, ApJ Letter, 644, L63

\bibitem[2002]{levinson02} 
Levinson, A., Ofek, E.~O., Waxman, E., \& Gal-Yam, A.\ 2002, \apj, 576, 923 

\bibitem[2004]{magnier}
Magnier, E. A., Cuillandre, J-C., 2004, PASP, 116, 449

\bibitem[1997]{meszaros97} 
M{\'e}sz{\'a}ros, P., \& Rees, M.~J.\ 1997, \apj, 476, 232 

\bibitem[2000]{moderski00} 
Moderski, R., Sikora, M., \& Bulik, T.\ 2000, \apj, 529, 151 

\bibitem[1998]{monet98} 
Monet, D.~B.~A., Canzian, B., Dahn, C. et al.\ 1998, VizieR Online Data Catalog, 1252, 0 

\bibitem[2002]{nakar02}
Nakar, A., Piran, T., \& Granot, J. 2002, ApJ, 579, 699

\bibitem[2001]{panaitescu01} 
Panaitescu, A., \& Kumar, P.\ 2001, \apj, 554, 667 

\bibitem[1998]{perna98} 
Perna, R., \& Loeb, A.\ 1998, \apjl, 509, L85 

\bibitem[2001]{piran01} 
Piran, T., Kumar, P., Panaitescu, A., \& Piro, L.\ 2001, \apjl, 560, L167 

\bibitem[2006]{rau}
Rau et al., 2006, A\&A, 449, 79

\bibitem[1992]{rees92} 
Rees, M.~J., \& M{\'e}sz{\'a}ros, P.\ 1992, \mnras, 258, 41P 

\bibitem[1994]{rees94} 
Rees, M.~J., \& M{\'e}sz{\'a}ros, P.\ 1994, \apjl, 430, L93 

\bibitem[1997]{rhoads97} 
Rhoads, J.~E.\ 1997, \apjl, 487, L1 

\bibitem[1999]{rhoads99}
Rhoads, J. E. 1999, ApJ, 525, 737

\bibitem[2005]{rykoff05}
Rykoff, E. S., Aharonian, F., Akerlof, C. W. et al. 2005, ApJ, 631, 1032

\bibitem[1998]{sari98} 
Sari, R., Piran, T., \& Narayan, R.\ 1998, \apjl, 497, L17 

\bibitem[1999]{sari99} 
Sari, R., Piran, T., \& Halpern, J.~P.\ 1999, \apjl, 519, L17 

\bibitem[2002]{totani02}
Totani, T., \& Panaitescu, A. 2002, ApJ, 576, 120

\bibitem[1995]{focas}
Valdes, F. G. et al. 1995, PASP, 107, 1119

\bibitem[2002]{van}
Vanden Berk, D. E. et al, 2002, ApJ, 576, 673

\bibitem[2000]{wei00} 
Wei, D.~M., \& Lu, T.\ 2000, \apj, 541, 203 

\bibitem[1997]{wijers97} 
Wijers, R.~A.~M.~J., Rees, M.~J., \& M{\'e}sz{\'a}ros, P.\ 1997, \mnras, 288, L51 

\bibitem[2005]{zeh}
Zeh, A., Klose, S. \& Kann, D. A. 2005, ApJ, 637, 889

\bibitem[2002]{zhang02} 
Zhang, B., \& M{\'e}sz{\'a}ros, P.\ 2002, \apj, 571, 876 

\end{thebibliography}
\end{document}